\documentclass[twocolumn,prl,showpacs,superscriptaddress]{revtex4}
\usepackage{graphicx}

\begin{document}
\title{Excitation hierarchy of the quantum sine-Gordon
spin chain in strong magnetic field}
\author{S. A. Zvyagin}
\affiliation{National High Magnetic Field Laboratory, Florida
State University, Tallahassee, FL 32310}
\author{A. K. Kolezhuk}
\affiliation{Institute of
Magnetism, National Academy of Sciences,  03142 Kiev, Ukraine}
\affiliation{Institut f\"{u}r Theoretische Physik, Universit\"{a}t
Hannover, 30167 Hannover, Germany}
\author{J. Krzystek}
\affiliation{National High Magnetic Field Laboratory, Florida
State University, Tallahassee, FL 32310}
\author{R. Feyerherm}
\affiliation{Hahn-Meitner-Institute (HMI), 14109 Berlin, Germany}

\begin{abstract}
The magnetic excitation spectrum of copper pyrimidine dinitrate, a
material containing $S=\frac{1}{2}$ antiferromagnetic chains with
alternating $g$-tensor and the Dzyaloshinskii-Moriya interaction,
and exhibiting a field-induced spin gap, is probed using
submillimeter wave electron spin resonance spectroscopy. Ten
excitation modes are resolved in the low-temperature spectrum, and
their frequency-field diagram is systematically studied in
magnetic fields up to 25 T. The experimental data are sufficiently
detailed to make a very accurate comparison with predictions based
on the quantum sine-Gordon field theory. Signatures of three
breather branches and a soliton, as well as those of several
multi-particle excitation modes are identified.

\end{abstract}
\pacs{75.40.Gb, 76.30.-v, 75.10.Jm}

\maketitle

\emph{Introduction.---} Recently, low-dimensional spin systems
have received a considerable amount of attention. This is
particularly due to their relevance to the quantum criticality
problem, which appears to be one of the central concepts in modern
solid state physics. Quantum fluctuations, significantly enhanced
in spin systems with reduced dimensionality, give rise to a
variety of strongly correlated states and make low-dimensional
magnets an ideal ground for testing various theoretical concepts.
To comprehend the role of quantum fluctuations in strongly
correlated electron and spin systems with reduced dimensionality,
it is important to explore its phenomenology in simple and
well-controlled model systems. In this context, understanding  the
nature of the ground state and excitations in quantum spin chains
is an important challenge. An isotropic $S=\frac{1}{2}$ Heisenberg
antiferromagnetic (AF) chain with uniform nearest-neighbor
exchange coupling represents one of the paradigm models of quantum
magnetism. Its ground state is a spin singlet, and the dynamics
are determined by a gapless two-particle continuum of
spin-$\frac{1}{2}$ excitations, commonly referred to as spinons. A
uniform external magnetic field causes a substantial rearrangement
of the excitation spectrum, making the soft modes incommensurate
\cite{Mueller,Stone}, although the spinon continuum remains
gapless. Since the $S=\frac{1}{2}$ AF chain is critical, even
small perturbations can considerably change fundamental properties
of the system. One of the most prominent examples is the
$S=\frac{1}{2}$ AF chain perturbed by an alternating $g$-tensor
and/or the Dzyaloshinskii-Moriya (DM) interaction; this situation
 is realized experimentally in several
spin chain systems \cite{Dender,Asano,Feyerherm,Kohgi,Kenzelmann}.
In the presence of such interactions, application of a  uniform
external field $H$ induces an effective transverse staggered field
$h\propto H$, which leads to the opening of an energy gap
$\Delta\propto H^{2/3}$. The gapped phase can be effectively
described by the quantum sine-Gordon field theory
\cite{OshikawaAffleck97,AffleckOshikawa99prb}. The  excitation
spectrum is represented by solitons, antisolitons, and multiple
soliton-antisoliton bound states called breathers. The
availability of exact solutions for the sine-Gordon model allows
very precise theoretical description of many observable properties
and physical parameters of sine-Gordon magnets, including field
dependence of excitation energies
\cite{OshikawaAffleck97,Essler99,AffleckOshikawa99prb} and
response functions \cite{Essler+,EsslerFH}. This makes such
systems a particularly interesting target for experimentally
probing elementary excitations.  So far, a field-induced gap has
been observed in several $S={1\over2}$ AF chain materials
\cite{Dender,Asano,Asano2,Feyerherm,Kohgi,Kenzelmann}. At the same
time, experimental information on the magnetic field behavior of
the spectrum is rather limited, and usually only the lowest one or
two excitations can be identified. In the most detailed existing
study \cite{Kenzelmann}, a soliton and two breather modes were
observed in CuCl$_2\cdot 2$(dimethylsulfoxide) (CDC) using
inelastic neutron scattering. However, significant interchain
interactions present in CDC lead to a pronounced deviation from
the  sine-Gordon model predictions. In ESR studies of copper
benzoate \cite{Asano,Asano2}, several modes were observed but only
the lowest one was identified.

 In this Letter we report a detailed study of the elementary
excitation spectrum in copper pyrimidine dinitrate (hereafter
Cu-PM), which has been recently identified  as a $S=\frac{1}{2}$
AF chain with a field-induced spin gap \cite{Feyerherm}, and is
probably the best realization of the quantum sine-Gordon spin
chain model known to date. Its intrachain exchange constant
$J=36$~K \cite{Feyerherm,Wolter+03} is a factor of four larger
than that in CDC \cite{Landee} and about two times larger compared
to copper benzoate \cite{Dender}. It makes Cu-PM an excellent
object for the experimental study of $S=\frac{1}{2}$ AF chains in
the low-temperature ``nonperturbative'' regime
\cite{OshikawaAffleck-esr}, where the temperature $T$ is small
compared to the energy gap $\Delta$.  By employing high-resolution
tunable-frequency submillimeter wave electron spin resonance (ESR)
spectroscopy, \emph{ten} different ESR modes were resolved and
their behavior in a wide range of magnetic fields up to
$g\mu_{B}H\sim J$ was studied.  By comparing the experimental data
with the calculations based on the quantum sine-Gordon field
theory, we were able to identify signatures of the \emph{three
lowest breathers} and of a \emph{soliton}.

\begin{figure}
\begin{center}
\vspace{1.5cm}
\includegraphics[width=0.48\textwidth]{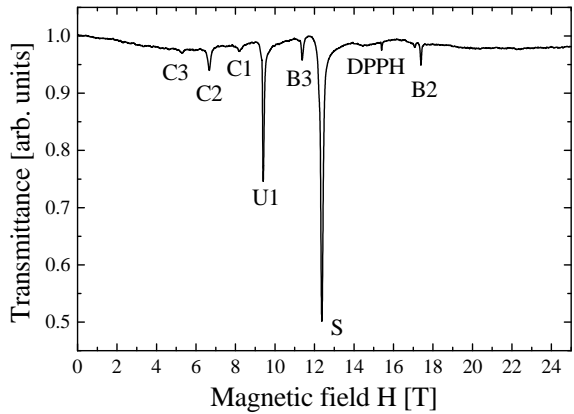}
\vspace{-2.3 cm} \caption{\label{fig:sample} The ESR transmission
spectra in Cu-PM, taken in the Faraday configuration at a
frequency of $429.3$~GHz at $T=1.6$~K (for explanations see the
text). DPPH was used as a marker.}
\end{center}
\end{figure}

\emph{Experimental.--- }
Cu-PM, [PM-$\rm Cu(NO_{3})_{2}(H_{2}O)_{2}$]$_{n}$
 (PM = pyrimidine)
crystallizes in a monoclinic structure belonging to space group
$C2/c$ with four formula units per unit cell \cite{Feyerherm}. The
lattice constants obtained from the single-crystal X-ray
diffraction are $a=12.404$~\AA, $b=11.511$~\AA, $c=7.518$~\AA,
$\beta=115.0^{\circ}$. The Cu ions form  chains (with a distance
$d=5.71$~\AA~  between neighboring $\rm Cu^{2+}$ ions  at 10 K)
running parallel to the short $ac$ diagonal. The Cu ions are
linked by the N-C-N moieties of pyrimidine, which constitute the
intrachain magnetic exchange pathway. The interchain Cu-Cu
distance is 6.84~\AA. The Cu coordination is a distorted
octahedron, built from an almost square N-O-N-O equatorial plane
and two oxygens in the axial positions. In this approximately
tetragonal local symmetry, the local principal axis of each
octahedron is tilted from the $ac$ plane by $\pm 29.4^{\circ}$.
Since this axis almost coincides with the principal axis of the
$g$-tensor,   the $g$-tensors for neighboring Cu ions are
staggered. The exchange constant $J=36\pm0.5$~K was extracted from
the single-crystal susceptibility \cite{Feyerherm} and confirmed
by magnetization measurements \cite{Wolter+03}.

The excitation spectrum was studied using a high-field
submillimeter wave ESR spectrometer \cite{Zvyagin}. Backward Wave
Oscillators were employed as tunable sources of radiation,
quasi-continuously covering the frequency range of 150 to 700 GHz.
These radiation sources in combination with the highly-homogeneous
magnetic field provided by a 25 T Bitter-type resistive magnet
make the facility a powerful tool for studying magnetic excitation
spectra in highly correlated spin systems.  The magnetic field was
applied along the $c''$ direction, which is characterized by the
maximal value of the staggered magnetization for Cu-PM
\cite{Feyerherm,Asano1}. Our measurements at room temperature
yield $g=2.24\pm 0.02$ as the effective $g$-factor for the field
applied along the $c''$ direction, which is consistent  with the
data of Ref.\ \cite{Feyerherm}. High-quality single-crystals of
Cu-PM were probed using both Faraday
 and Voigt geometry sampleholders.
 We found that the spectra obtained in the
Faraday and Voigt configurations exhibit a pronounced similarity
(possible reasons for that will be discussed below).

\begin{widetext}

\begin{figure}[tb]
\begin{center}
\includegraphics[width=0.82\textwidth]{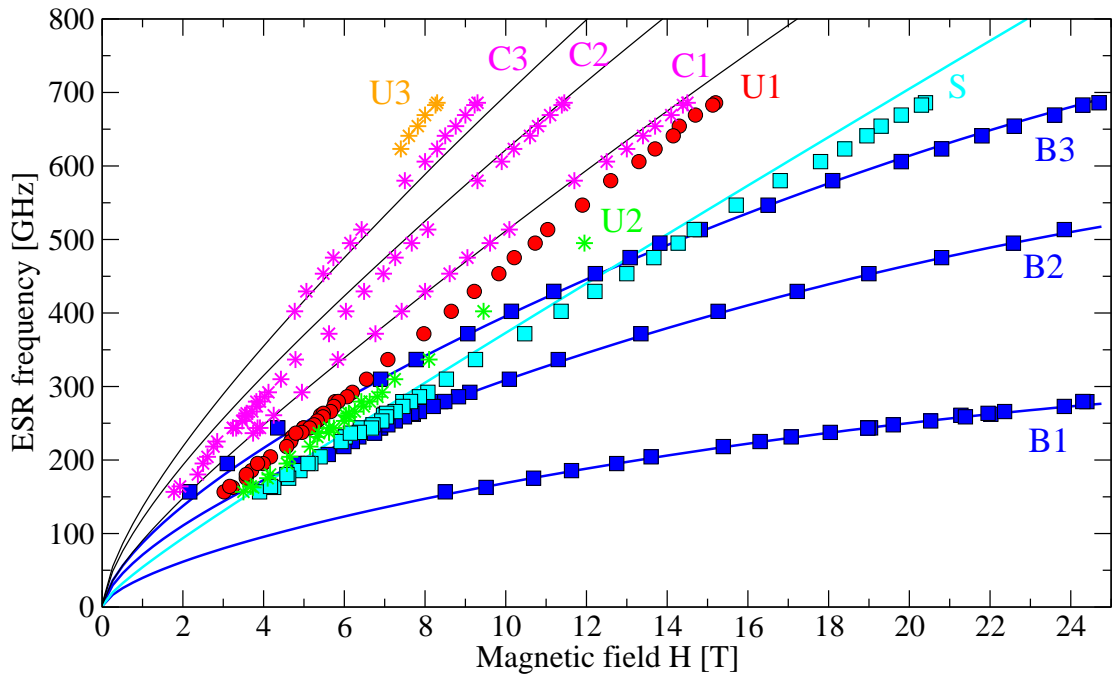}
\caption{\label{fig:FFD} The frequency-field dependence of the ESR
modes in Cu-PM at $T=1.6$~K. Symbols denote the experimental
results, and lines correspond to contributions from specific
excitations as predicted by the sine-Gordon theory (see
the text).}

\end{center}
\end{figure}

\end{widetext}

Several resonance modes with different intensities were observed
in experiments. A typical ESR transmittance spectrum obtained at a
frequency of 429.3 GHz and at temperature $T=1.6$ K is shown in
Fig.\ \ref{fig:sample}. Within the experimental accuracy, the
absorptions can be nicely fit using the Lorentzian formula for the
line shape. The complete frequency-field diagram of magnetic
excitations, collected at $T=1.6$ K, is presented in Fig.\
\ref{fig:FFD}. Absorptions denoted as  $\rm B1$, $\rm S$ and  $\rm
U1$ have maximal intensity, while the rest of excitations are
about one order of magnitude weaker. It is worthwhile to mention
here that the  low-temperature behavior  of their integrated
intensity, carefully checked at several temperatures down to 1.6
K,  strongly suggests the ground state nature of the observed
excitations (details of the temperature evolution of the
excitation spectrum in Cu-PM will be reported elsewhere
\cite{Zvyagin2}).

\emph{Discussion.---} The experimental frequency-field diagram has
been analyzed in the framework of the quantum sine-Gordon field
theory approach \cite{OshikawaAffleck97,AffleckOshikawa99prb}.  We
used the following expression \cite{AffleckOshikawa99prb} for the
soliton gap $\Delta_{s}$ which is valid for a wide range of fields
up to $g\mu_{B}H\sim J$:
\begin{equation}
\label{solgap}
\Delta_{s}=J {2\Gamma({\xi\over2})v_{F} \over
  \sqrt{\pi}\Gamma(\frac{1+\xi}{2})}
\left[ {g\mu_{B}H \over J v_{F}} {\pi\Gamma({1\over 1+\xi}) c A_{x} \over
  2\Gamma({\xi\over 1+\xi})}\right]^{\frac{1+\xi}{2}}.
\end{equation}
Here $c$ is the proportionality coefficient connecting the uniform
applied field $H$ and the effective staggered field $h=cH$, the
parameter $\xi=\big(2/(\pi R^{2})-1\big)^{-1}$, where $R$ is the so-called
compactification radius, and $v_{F}$ has the meaning of the Fermi
velocity. Both $R$ and $v_{F}$ are known exactly  as  functions of
$\widetilde{H}=g\mu_{B}H/J$ from the solutions of the Bethe ansatz
equations \cite{AffleckOshikawa99prb}. The amplitude $A_{x}$,
which is also a function of $\widetilde{H}$, was recently computed
numerically \cite{EsslerFH}.
At a given field $H$, there are $N=[1/\xi]$ breather
branches $B_{n}$ with $n=1,\ldots N$.
The breather gaps $\Delta_{n}$ are given by the
formula
\begin{equation}
\label{bregap} \Delta_{n}=2\Delta_{s}\sin(n\pi\xi/2). \quad
\end{equation}
At $H$=0 the first breather $B_{1}$ is degenerate with the
soliton-antisoliton doublet $S$, $\overline{S}$. At finite $H$
this degeneracy is lifted, so that $B_{1}$ becomes the lowest
excitation and gives the strongest contribution into the magnitude
of the gap observed in specific heat experiments \cite{Feyerherm}.
The sine-Gordon model predicts two more ``heavy'' breathers
$B_{2}$, $B_{3}$ to exist in the relevant frequency-field range.

ESR probes the dynamical susceptibility $\chi(q,\omega)$ at the
momentum $q=0$. However, in the present case the alternation of
the $g$-tensor and DM interaction leads to a mixing of $q=0$ and
$q=\pi$ components. Moreover, due to the effective redefinition of
spin operators which is performed to eliminate the DM interaction
\cite{AffleckOshikawa99prb}, and due to the fact that the
Dzyaloshinskii vector is directed at an angle of about
$58^{\circ}$ with respect to the magnetic field \cite{Feyerherm},
the physical susceptibility $\chi_{\rm phys}$ is generally a
mixture of effective longitudinal and transverse susceptibilities
$\chi_{sG}$ calculated within the sine-Gordon model. For instance,
the physical transverse susceptibility  $\chi_{\rm
phys}^{\perp}(0,\omega)$ is determined not only by
$\chi_{sG}^{\perp}(0,\omega)$, but also gets a finite contribution
from $\chi_{sG}^{\perp}(\pi,\omega)$ and
$\chi_{sG}^{\parallel}(\pi,\omega)$;  similarly, the longitudinal
component $\chi_{\rm phys}^{zz}(0,\omega)$ contains a small
admixture of $\chi_{sG}^{xx}(0,\omega)$ and
$\chi_{sG}^{yy}(\pi,\omega)$. For the analysis of inelastic
neutron experiments probing $q\approx\pi$, this mixing could be
safely neglected \cite{AffleckOshikawa99prb} since typically
$\chi(0,\omega)\ll\chi(\pi,\omega)$, but it becomes important for
the interpretation of ESR spectra, where even a small admixture of
$\chi(\pi,\omega)$ can have intensity comparable to that of
$\chi(0,\omega)$. As a result, one should be able to observe
contributions from $\chi_{sG}^{\perp}$ as well as from
$\chi_{sG}^{\parallel}$ both in the Faraday and Voigt geometries.

\begin{figure}[tb]
\begin{center}
\includegraphics[width=0.4\textwidth]{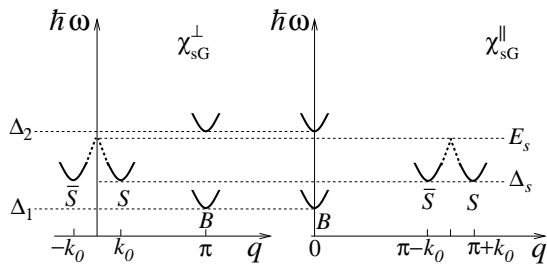}
\caption{\label{fig:scheme} A schematic view of low-energy
excitations contributing to the transverse (left panel) and
longitudinal (right panel) dynamic susceptibilities;
$\overline{S}$ ($S$) denotes (anti)solitons and $B$ labels
breathers.
}
\end{center}
\end{figure}

The energy structure of elementary excitations of the sine-Gordon
model contributing to the low-energy spin dynamics is sketched in
Fig.\ \ref{fig:scheme}.  Single-particle contributions to the
longitudinal susceptibility $\chi_{sG}^{\parallel}$ are determined
by solitons concentrated around incommensurate wave vectors
$q=\pi\pm k_{0}$ and by breathers concentrated around $q=0$. Here
the incommensurate shift $k_{0}=2\pi m$ is determined by the total
magnetization per spin $m$, exactly known as a function of the
field; for small $H$ one has $J v_{F}k_{0}\approx g\mu_{B}H$.  In
the transverse susceptibility $\chi_{sG}^{\perp}$ the dominating
contribution comes from breathers at  $q=\pi$ and solitons at
$q=\pm k_{0}$.  Thus, in an ESR experiment there should be several
breather resonances at the energies $\Delta_{n}$ as well as a
single soliton resonance at
\begin{equation}
\label{solD} E_{s}\simeq\sqrt{\Delta_{s}^{2}
  +(Jv_{F}k_{0})^{2}}.
\end{equation}
This exhausts the set of possible single-particle resonances (see
Fig.\ \ref{fig:scheme}); apart from them, there are various
multiparticle continua contributing to the spectrum.

It is important to mention that in Eqs.\
(\ref{solgap})-(\ref{solD}) \emph{the~only} free fitting parameter
is the coefficient $c$. Setting $c=0.08\pm 0.002$,  we were able
to achieve an excellent fit to the lowest observed mode $\rm B1$,
which is described by the first breather gap $\Delta_{1}$. Using
this value of $c$, we calculated energies of other modes predicted
by the sine-Gordon model, and obtained a reasonably good fit to
the entire set of the experimental data. On the basis of the fit,
we identify the observed resonances as follows: the modes
$\rm B1$, $\rm B2$ and $\rm B3$ correspond to the first three breather resonances at
$\Delta_{1}$, $\Delta_{2}$ and $\Delta_{3}$, respectively
(blue lines in
Fig.\ \ref{fig:FFD}). The mode $\rm S$ is well fitted by the soliton
resonance at $E_{s}$ (the cyan line in Fig.\ \ref{fig:FFD}). This
interpretation is supported by
the analysis of the temperature dependence of the ESR spectra,
which shows that the $\rm S$ mode
continuously transforms into the $\omega=g H$ resonance when the
temperature increases, in agreement with the theory \cite{OshikawaAffleck-esr}.
The overall agreement between the sine-Gordon theory predictions
and the experimentally obtained frequency-field dependence for those
four single-particle resonances is very
good. Note however that our value $c=0.08$ differs from the result
$c=0.11$ obtained recently \cite{Wolter+03} from the analysis of
magnetization curves of Cu-PM; the origin of this discrepancy is  not clear at present.

 The identification of the other
six high-frequency ESR modes is more difficult since the theory
does not predict any single-particle contributions in the relevant
frequency region. One may nevertheless speculate that modes $\rm
C1$-$\rm C3$ lie very close to the edges of the soliton-breather
continua $SB_{1}$, $SB_{2}$ and $SB_{3}$, respectively, denoted by
black lines in Fig.\ \ref{fig:FFD}. The edges of the $B_{1}B_{1}$
and $S\overline{S}$  continua were not observed, presumably
because they are close to the lines $\rm B1$ and $\rm B2$ and are
masked by the corresponding quasiparticle contributions.   For the
remaining three modes $\rm U1$-$\rm U3$ we were not able to find
any appropriate explanation on the basis of the sine-Gordon model,
although we checked all possible two- and three-particle continua.
This is especially puzzling in case of the $\rm U1$ mode, which is
one of the most intensive excitations (see Fig.\
\ref{fig:sample}). We hope that our observations will stimulate
further theoretical work in this direction.

In summary, we have presented a detailed frequency-field diagram of
spin excitations in Cu-PM, a material containing $S=\frac{1}{2}$ AF
chains with alternating $g$-tensor and DM interaction and exhibiting a
field-induced gap.  The use of high-resolution submillimeter wave ESR
spectroscopy made possible obtaining a very precise information on the
magnetic excitation spectrum.  The field-induced gap was observed
\emph{directly}, and its high-field behavior was studied. \emph{Ten}
ESR modes were resolved and their behavior in a broad field range up
to $g\mu_{B}H\sim J$ was studied.  By comparing the entire set of data
with the theoretical predictions, we have provided experimental
evidence for a number of excitations of the sine-Gordon theory,
including \emph{solitons} and the \emph{three} lowest members of the
breather hierarchy.  At the same time, we observed at least one strong
high-frequency resonance mode which does not fit into the sine-Gordon
model description.  Our results can be relevant to understanding the
quantum spin dynamics in copper benzoate and other $S=\frac{1}{2}$
Heisenberg antiferromagnetic chain systems.

\emph{Acknowledgments.--} We express our special thanks to
F.~H.~L.~Essler and C.~Broholm for critical reading of the
manuscript and valuable comments. We are also grateful to
A.~Honecker and A.~U.~B.~Wolter for fruitful discussions.
Experiments performed at NHMFL were supported by the NSF through
Cooperative Grant DMR-9016241. A.K. was supported by Grant I/75895
from Volkswagen-Stiftung.

\end{document}